\documentclass[a4paper,fleqn,usenatbib]{mnras}

\usepackage{newtxtext,newtxmath}

\usepackage{adjustbox}

\usepackage[skip=0.333\baselineskip]{caption}
\usepackage[T1]{fontenc}
\usepackage{ae,aecompl}
\usepackage{hyperref}
\usepackage{natbib}
\usepackage{listings}
\usepackage{graphicx}
\usepackage{xspace}	
\usepackage{amsmath}
\usepackage{xparse}
\usepackage{parskip}

\title[The Reflectance Spectra of CCs]{
The Reflectance Spectra of CV-CK Carbonaceous chondrites from the Near Infrared to the Visible
}

\author[S. Tanbakouei et al. ]{
S. Tanbakouei ,$^{1,2}$\thanks{E-mails: tanbakouei@ice.csic.es, trigo@ice.csic.es}
Josep M. Trigo-Rodríguez ,$^{1,2}$
J. Llorca,$^{3}$ \newauthor
C.E. Moyano-Cambero,$^{1,2}$
 I.P. Williams,$^{4}$
Andrew S. Rivkin $^{5}$
\\
$^{1}$Institute of Space Sciences (CSIC), Meteorites, Minor Bodies and Planetary Sciences Group, Campus UAB, c/Can Magrans s/n, \\08193 Cerdanyola del Vallès (Barcelona), Catalonia, Spain\\
$^{2}$Institut d’Estudis Espacials de Catalunya (IEEC), Gran Capità, 2-4, despatx 201, 08034 Barcelona, Catalonia, Spain\\
$^{3}$Institute of Energy Technologies, Department of Chemical Engineering and Barcelona Research Center in Multiscale Science and Engineering,\\ Universitat Politècnica de Catalunya, EEBE, 08019 Barcelona, Catalonia, Spain\\
$^{4}$School of Physics and Astronomy, Queen Mary, University of London, Mile End Rd. London E1 4NS, UK.\\
$^{5}$John Hopkins University Applied Physics Laboratory, Laurel, MD, USA.\\
}
\date{Accepted XXX. Received YYY; in original form ZZZ}

\pubyear{2019}

\usepackage{xparse}
\ExplSyntaxOn
\seq_new:N \g_yasamin_boldcite_seq
\cs_new_eq:NN \__yasamin_org_bibitem:wn \bibitem
\RenewDocumentCommand \bibitem { O{} m }
  {
  \seq_if_in:NnTF \g_yasamin_boldcite_seq {#2}
     { \__yasamin_boldify_bibitem:nnw {#1} {#2} }
      { \__yasamin_org_bibitem:wn [#1] {#2} }
 }
\cs_new_protected:Npn \__yasamin_boldify_bibitem:nnw #1 #2 #3 \par
  {
    \__yasamin_org_bibitem:wn [#1] {#2}
    \group_begin:
     \bfseries #3 \tex_par:D
    \group_end:
 }
\NewDocumentCommand \makeboldref { m }
{
  \seq_set_from_clist:Nn \l_tmpa_seq {#1}
   \seq_gconcat:NNN \g_yasamin_boldcite_seq \g_yasamin_boldcite_seq \l_tmpa_seq
  }
\ExplSyntaxOff

\begin{document}
\label{firstpage}
\pagerange{\pageref{firstpage}--\pageref{lastpage}}
\maketitle
\begin{abstract}
Carbonaceous chondrite meteorites are so far the only available samples representing carbon-rich asteroids and in order to allow future comparison with samples returned by missions such as Hayabusa 2 and OSIRIS-Rex, is important to understand their physical properties. Future characterization of asteroid primitive classes, some of them targeted by sample-return missions, requires a better understanding of their mineralogy, the consequences of the exposure to space weathering, and how both affect the reflectance behavior of these objects. In this paper, the reflectance spectra of two chemically-related carbonaceous chondrites groups, precisely the Vigrano (CVs) and Karoonda (CKs), are measured and compared. The available sample suite includes polished sections exhibiting different petrologic types: from 3 (very low degree of thermal metamorphism) to 5 (high degree of thermal metamorphism). We found that the reflective properties and the comparison with the Cg asteroid reflectance class point toward a common chondritic reservoir from which the CV-CK asteroids collisionally evolved. In that scenario the CV and CK chondrites could be originated from 221 Eos asteroid family, but because of its collisional disruption, both chondrite groups evolved separately, experiencing different stages of thermal metamorphism, annealing and space weathering. 
\\
\end{abstract}


\begin{keywords}
\\
Asteroids-Meteoites-Minor bodies
\end{keywords}

Accepted for publication in MNRAS on July 19th, 2021 



\section{Introduction}
Most meteorites originates from asteroids and as such they provide samples that can be directly studied in laboratories. Though they have different size scales, an asteroid and a meteorite can share compositional and reflectance properties because they contain similar rock-forming minerals \citep{cloutis2012spectral, trigo2015orbit}. The undifferentiated chondritic meteorites have proven to be good proxies to understand the properties of chondritic asteroids \citep{cloutis2012spectral, trigo2014uv, beck2014transmission, moyano2017nanoindenting, tanbakouei2019mechanical}. Meteorites can be studied  in through dedicated laboratory experiments,  but the properties of an asteroid are usually inferred from ground-based observations, though a few also have  been observed from spacecraft \citep[see e.g.][] {gaffey2002asteroids}. Asteroids are subjected to space weathering processes that alter the initial surface properties of the body \citep{brunetto2015asteroid, colonna2019hypersonic}. Collisional evolution of asteroids provides meter-sized meteoroids that are producing meteorite falls \citep{trigo2009tensile, beitz2016collisional}.  So that, the Earth receives a continuous delivery of asteroidal materials providing invaluable source of scientific information \citep{colonna2019hypersonic}. Each one of these rocks has the potential to explain the story of its progenitor body and provide information about the processes that occurred in their parent asteroids \citep{binzel2010earth, binzel2015near}. For instance, collisional gardening has been at work for 4.5 Gyrs and this bombardment modified the primordial properties of chondritic asteroids \citep{beitz2016collisional, blum2006physics}. The spectral properties of the near-Earth target asteroid 162173 Ryugu reported by Hayabusa 2 mission, have compared so far by \citet{kitazato2019surface} with the meteorites of CMs, CIs and CV3  Allende. Another mission, called OSIRIS - REx (Origins, Spectral Interpretation, Resource Identification, Security, Regolith Explorer), explored the near-Earth asteroid (101955) Bennu providing a good insight about the reflectance properties of the asteroid in the visible and near infrared range \citep{lauretta2017osiris, barucci2020osiris, rizos2021bennu}. \citet{hamilton2019evidence} also established a comparison with CMs and CIs chondrites to infer the
different degrees of hydration of the parent asteroids.

Laboratory reflectance spectra of meteorites are often used to identify the main rock-forming minerals of asteroids \citep{gaffey2002asteroids, cloutis2011spectral, cloutis2012spectral, beck2014transmission}. Over the years, additional information such as physico-chemical properties of carbonaceous chondrites (CCs) and primitive asteroids and their reflective behavior, was obtained by increasingly sensitive and precise spectrometers \citep{barucci2005asteroid, cloutis2011spectral, trigo2014uv}. Carbonaceous chondrites come from undifferentiated objects with sizes between one and hundreds of kilometres, formed in the outer region of the main asteroid belt (MB) or beyond, and sculpted by collisions \citep{chapman1975surface, vernazza2011asteroid, beitz2016collisional}. The characterization of carbonaceous chondrites  is often challenging because their reflectance spectra are dark, sometimes involving silicates and oxides, but also including minor contribution from opaque amorphous materials. The crystalline and amorphous materials within protoplanetary disks are thermally processed and accreted to make bodies of various sizes and compositions, depending on the heliocentric distance of formation \citep{trigo2019accretion}. The presence of some of these components alter their light scattering properties, and link these meteorites to some of the darkest objects of our Solar System \citep{cloutis2011spectral}. 

In this paper we study the similarities in the reflectance behavior of the CV group- named after Vigarano meteorite- and the CK group, named after Karoonda, given that a chemical similarity was already established between both groups previously \citep{greenwood2010relationship, wasson2013compositional}. According to the relation between CV and CKs, different degrees of thermal metamorphism and aqueous alteration, resulting from the asteroidal collision and differentiated metamorphism could be studied \citep*{wasson2013compositional, trigo2015aqueous, chaumard2016chondrules, rubin2017meteoritic}. 

We explore a scenario in which CV and CK chondrites evolved differently as consequence of the catastrophic disruption of a moderately large progenitor asteroid. The surviving asteroid fragments were dynamically separated and exposed to impacts with other bodies, so at the end the degree of oxidation was different. Oxidation, experienced by the presence of Fe$^{+3}$ in low-Ca pyroxenes, and abundant Ni in olivine, are indicative that metamorphism occurred in most CK samples, since petrographic type 4 or higher types indicate thermal metamorphism \citep*{noguchi1993petrology}. Such physical process also leads to Fe enrichment in CAIs \citep{chaumard2014metamorphosed}. As a consequence of metamorphism, the total carbon contents of CKs tend to decrease with increasing metamorphic grade \citep{jarosewich2006chemical, greenwood2010relationship}. In fact, \citet*{geiger1995formation} suggested that CK chondrites formed by metamorphic heating of a CV like precursor under oxidizing conditions. Some part of the metamorphism is due to collisions and the associated shocks, and so the resultant minerals are those being identified in our spectra \citep*{urzaiz2015identification}. \citet{huber2006siderophile} chose CK samples of un-weathered Antarctic meteorites to analyse the bulk elemental chemistry using INAA (Instrumental Neutron Activation Analysis) and petrographic techniques. These authors concluded that, as consequence of the aqueous alteration on the parent asteroid, CK chondrites became highly oxidized \citep{clayton1999oxygen}.

All CK chondrites contain shock-induced fractured and darkened silicates exhibiting tiny (< 0.3-10 $\mu$m) grains of magnetite and pentlandite throughout the interiors of many silicate grains \citep*{brearley1998reviews, kallemeyn1991compositional}. \citet*{hashiguchi2008mineralogy} proposed that these opaque regions were produced as a consequence of the progressive shock-darkening that creates vesicles and holes. The olivine in CK chondrites had been seen as spherical grains (0.1–5 ${\mu}$m) of magnetite and pentlandite, and other silicates. \citet*{noguchi1993petrology} and \citet*{ geiger1995formation} remarked that $\mu$m-sized magnetite grains act as opaque in the interior of silicates, thus contributing to shock darkening. The process of post-shock annealing could led to minimize influence of shock in olivine \citep*{bauer1979experimental, rubin2004postshock, rubin2012collisional}. According to \citet*{brearley1991subsolidus} silicate blackening could be due to shock but other processes could have been at work.

We use CV and CK meteorite reflectance spectra to identify specific spectral features that can be used to characterize the parent bodies of these meteorites in remote telescopic observations. As these groups have experienced different degrees of metamorphism, their reflectance behavior might help us to establish some evolutionary patterns with other CC classes, which we wish to identify how these patterns deal with the effects of shock in the history of their parental asteroids. Another clear example of the importance of the collisional history of asteroids is (2) Pallas \citep{marsset2020violent}. Establishing new reflectance spectra links is of key importance as collisionally processed transitional asteroids and short period comets could also produce carbonaceous chondrites \citep{tanbakouei2020comparing}.

The paper is organized as follow: In Section 2,  sample selection and the instrumental procedures are explanied in details. One section is also didecated to data analysis which explains the parameters in Table~\ref{tab3}. In Section 3 the results and obtained spectrcal reflectivity figures are provided. Section 4 will talk about the comparisons of our result with previuos methods, comparison of the reflectivity between CVs and Cks, and comparison of these chondrites with Cg asteroidal classes. Also a brief paragraph is provided about the Eos Family compare to CK carbonaceous chondrites. Section 5 is the conclusion part of the study. Appendix is provided as some supplmentary figures.

\section{Instrumental procedure}
\subsection{Samples characterization using microscope techniques}

At the Institute of Space Sciences meteorite clean laboratory, we previously study the sections using a Zeiss Scope, with magnifications up to 500X, and with a Motic BA310Pol Binocular microscope. Main minerals are identified using Reflected (RL) and Transmitted light (TL) techniques and later confirmed using  Energy-Dispersive X-ray Spectrometry (EDS) using an electron microscope (see below). In this way we recognize the principal features and minerals forming each rock (Table~\ref{tab1}). In addition, TL informs about the degree of terrestrial weathering experienced by each meteorite. Transmitted light (TL) microscopy is particularly useful as allows distinguishing the action of water in some aqueous alteration minerals like e.g. oxides and clays. Under the action of water crystalline silicates become blurry and show distinguishable alteration colors and textures. We always look for well-preserved sections, exhibiting a minimum terrestrial weathering. We applied similar methods for characterizing aqueous alteration as in chondrites \citep{trigo2006non, rubin2007progressive, trigo2015aqueous, trigo2019accretion} allow us to make sure that our thin sections are not too altered. Rusty interiors with Fe-oxides can be clearly identified as an orange background in transmitted light, not observed in our selected sections.

Once we identify a quite pristine section, we bring it for study using a Scanning Electron Microscope (SEM) at the Institut Català de Nanociència i Nanotecnologia (ICN2). In particular we used a FEI Quanta 650 FEG with a Back Scattered Electron Detector (BSED). This SEM was operated in low-vacuum mode with each thin section uncoated so we can detect the C abundance using a Energy-Dispersive X-ray Spectrometer (EDS) attached to the SEM. In particular, the instrument is an Inca 250 SSD XMax20 with Peltier cooling and with a detector area of 20 mm${^2}$. These analyses were used over selected 1 mm${^2}$ regions of interest (ROIs) and also on specific points in order to understand the sample’s mineralogy. Using all these chemical data we can rule out possible weathered areas. This is an important point as we wish having a minimum participation of minerals produced during the stay of meteorites on Earth.

\subsection{UV-NIR spectroscopy and data analysis}

The next step is to obtain the reflectance spectrum of each meteorite. To make sure that the reflectance spectrum is not affected by terrestrial weathering we chose a ROI of the interior of the meteorite section, far from the meteorite thin section. A Shimadzu UV3600 UV–Vis–NIR spectrometer was used to obtain reflectance spectra of thick and thin meteorite sections. As our spectrometer has a slot window of 1 $\times$ 0.2 mm${^2}$ we chose regions unaffected by weathering to be as much representative of the original parent body mineralogy.

Several CV and CK carbonaceous chondrites from the NASA Antarctic collection were selected (Table~\ref{tab1}) and measured using the spectrometer setup described by \citep{trigo2014uv}. The standard setting for the spectrometer is an integrating sphere (ISR) with a working range of 180-2600 nm, although the signal becomes noisy beyond about 2000 nm. In order to cover a wide range of wavelengths, the spectrometer uses multiple lamps, detectors and diffraction gratings which produce some artefacts that are removed in the spectra presented below.

We used thin ($\sim$30 $\mu$m) and thick ($\sim$1 mm) meteorite sections to obtain the spectra. All our samples had a larger area than the spectrometer beam to avoid signal reflections by other materials, following procedure \citep*{trigo2015aqueous}. \citet{moyano2016plausible} demonstrated using the same instrument that the incident light is not transmitted through the section. The spectra presented here covers the 400 to 1800 nm spectral range and allows comparison with asteroidal spectra taken from ground or space-based telescopes.

The spectrometer diffraction illumination originates from one of two lamps and passes through a variable slit, then is filtered with a grating to select the desired wavelength and afterwards is split into two alternating but identical beams with a chopper. Next the beam interacts with the sample and is routed to the detector. The reference beam interacts with the material and then goes to the same detector. The interior of the integrating sphere is coated with a duraflect reflecting polymer. The spectra obtained via this specific spectrometer, always shows noise between 800 and 900 nm plus two instrumental peaks (one from 1500 to 1600 nm and the other from 1900 to 2200 nm) \citep{moyano2016plausible}. Above 2000 nm, the spectra becomes noisy and loses uniformity, due mostly to humidity, carbon dioxide, and system hardware. To avoid displaying unreliable data, the spectra above the 1800 nm are not included. The peaks at 1500-1600 nm and at 1900-2200 nm are attributed to an unsuccessful instrumental correction applied by the software to remove the presence of the BaSO$_{4}$ (Fig~\ref{Fig1} in the appendix) substrate of the final spectra that has been previously explained in these papers {\citep{moyano2016plausible, tanbakouei2019mechanical}}. A new correction had been applied to the spectra to remove those peaks, but the final result in those regions became a bit noisier than desired. The region between 800 and 900 nm was deleted, as it does not include usable data and was too noisy to be reliable. The meteorite specimens studied here are listed in Table~\ref{tab1}. We have tried to chose samples from different petrologic types from CK3 to CK5.

\begin{table}
	\centering
\fontsize{10}{20}\selectfont
	\caption{Meteorite specimens analyzed, petrologic type, estimated weathering grade, oxidation state and main rock-forming minerals.}
	\label{tab1}
\begin{adjustbox}{center, width=\columnwidth-120pt}
	\begin{tabular}{c    c   c c c } 
		\hline
		Meteorite name &  Petrologic type & Weathering grade & Oxidation state & Dominant rock-forming minerals \\
		\hline
               
		Allende &	CV3 & W1 & Oxidized & Oliv., pyrox., troilite and CAIs \\
		ALH 84028	  & CV3 &  W0 & Oxidized & Oliv., pyrox., trolilite and CAIs\\
		ALH 85002    & CK4 & W1 & Oxidized & Oliv., pyrox., plagioclase and magnetite\\
          LAR 12265    & CK5 & W2 &  Oxidized &	Oliv., pyrox., sulphide and magnetite (shock darkened)\\
		LAR 04318 & CK4  	& W1 & Oxidized &  Oliv., pyrox., sulphide and magnetite\\
		MET 00430 & CV3	& W2 &  Oxidized  & Oliv., pyrox., and CAIs\\
		MET 01017 & CV3-an &W2 & Reduced & Oliv., pyrox., metal and sulphide \\
                     MET 01074 & CV3 & W1 &  Oxidized & Oliv., pyrox., and CAIs \\
                    MIL 07002 & CV3 & W2 &  Oxidized & Oliv., pyrox., and CAIs \\
                    PCA 82500 & CK3 & W2-3 &   Oxidized & Oliv., pyrox., troilite, metal and CAIs \\
\hline
	\end{tabular}
\end{adjustbox}
\end{table}

To obtain the spectral features of different minerals existed on our studied samples, the main peak of each mineral band is identified. The band center and width of the peaks are given in Table~\ref{tab3}. 
 All the spctral parameters including maximum and minimum peaks, their centers, and the width of each band are measured by a python code. The instrumental error is basically the step of the spectrometer which is $\sim$1 nm.

\section{Results}
\begin{figure}
\includegraphics[width=80mm, scale=0.7]{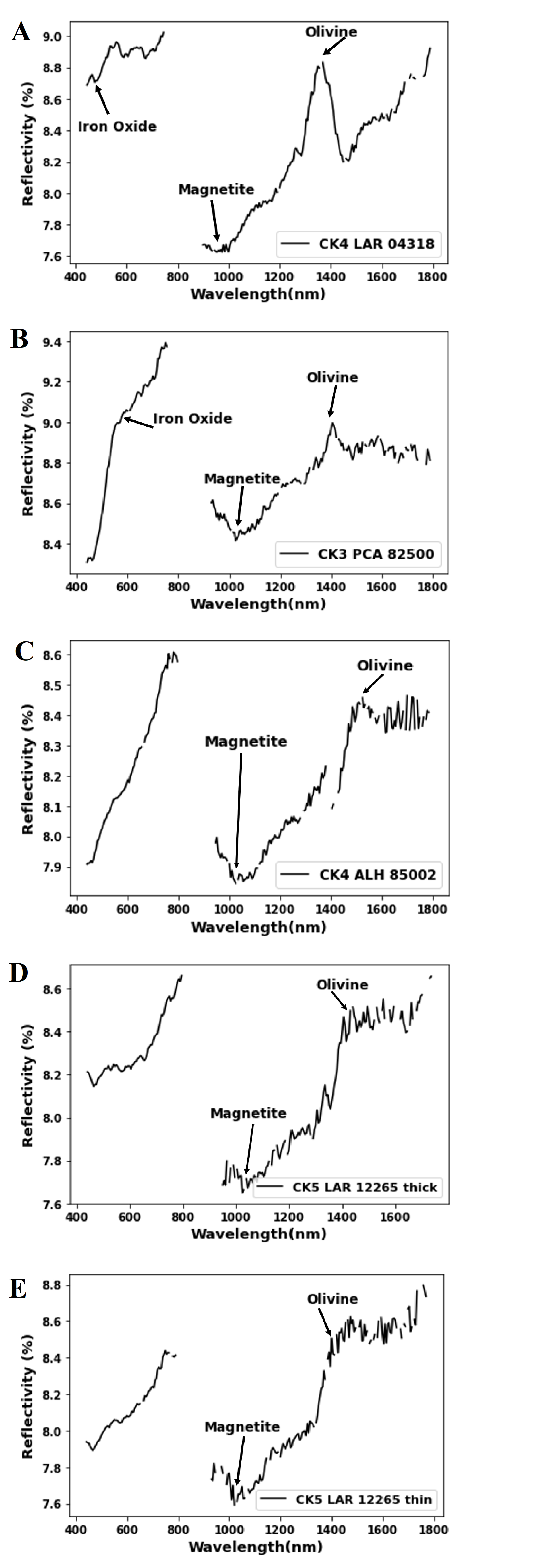}
    \caption{The reflectance spectra of five CK chondrites. The arrows mark the band positions of various minerals}.
    \label{fig1}
\end{figure}

The discovery of a significant number of new CK chondrites in Antarctica has increased the availability of such chondrites. Some of these new CK chondrites exhibit low levels of terrestrial weathering, making them a good source for spectroscopic comparison with other carbonaceous chondrites and to some extent with asteroids. The reflectance spectra of five CK chondrites obtained in this work are shown in Fig~\ref{fig1}. We used the average data among CK spectral reflectance and made the BaS$O_{4}$ correction above. Pecora Escarpment (PCA 82500) and Larkman Nunatak 04318 (LAR 04318) were corrected two times to reduce noises \citep{moyano2016plausible}.

The newly-obtained reflectance spectra of CK chondrites sections included in this study are shown in Fig~\ref{fig1} with different wavelength ranges. They have slightly red to blue spectral slopes beyond 900 nm (mostly red in the 900 to 1800 nm region and a bit blue slopped, above 1500 nm), exhibiting an olivine-like absorption band in the 1000-1200 nm region ( Fig~\ref{fig1}).

\begin{figure}
\includegraphics[width=80mm, scale=1.0]{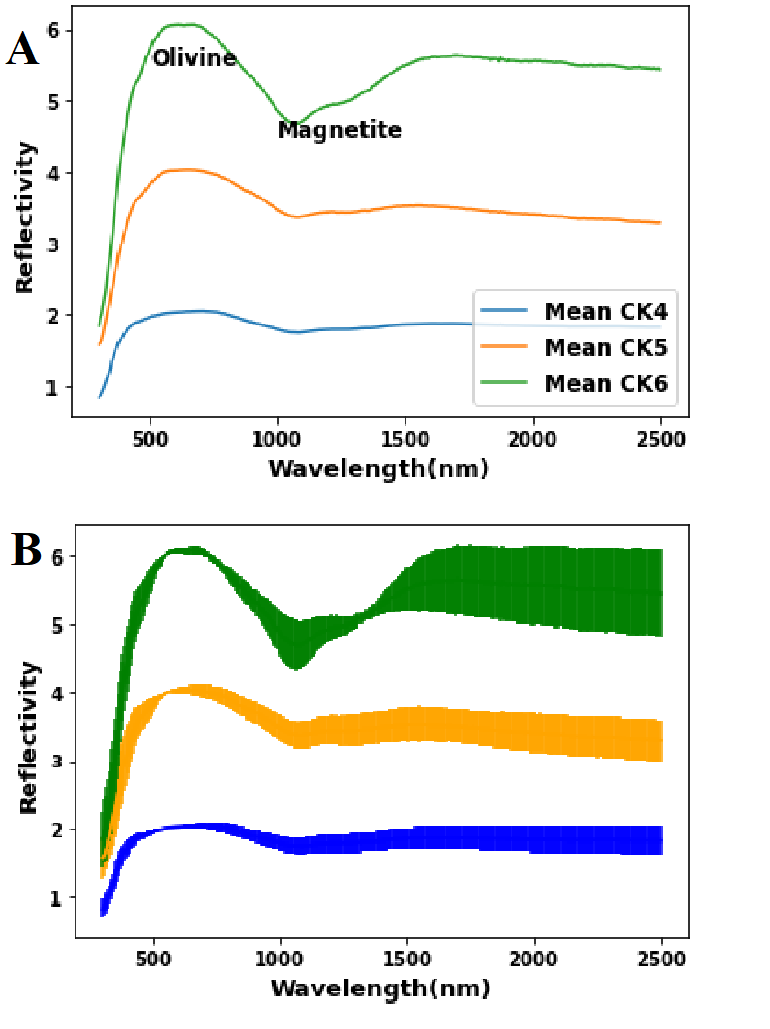}
    \caption{Comparison of CK reflectance spectra from meteorite powders studied in the RELAB spectral database in the range of 300 nm to 2500 nm range. a) Mean spectral reflectance of three different petrological types of CKs. b) Mean spectral reflectance of three different petrological type of CKs with error bars.}
    \label{Fig2}
\end{figure}

In the CK spectra it can be seen that the centre of the absorption bands does not change its position with the metamorphic grades, but correlates with the olivine composition. The depth of the bands, however, increases with higher degrees of thermal metamorphism, with the deepest band being found in CK6 spectra (Fig~\ref{Fig2}),  while CK3–5 petrographic grades display shallower band depths (Fig~\ref{Fig2}).

\begin{figure}
	\includegraphics[width=95mm, scale=2.5]{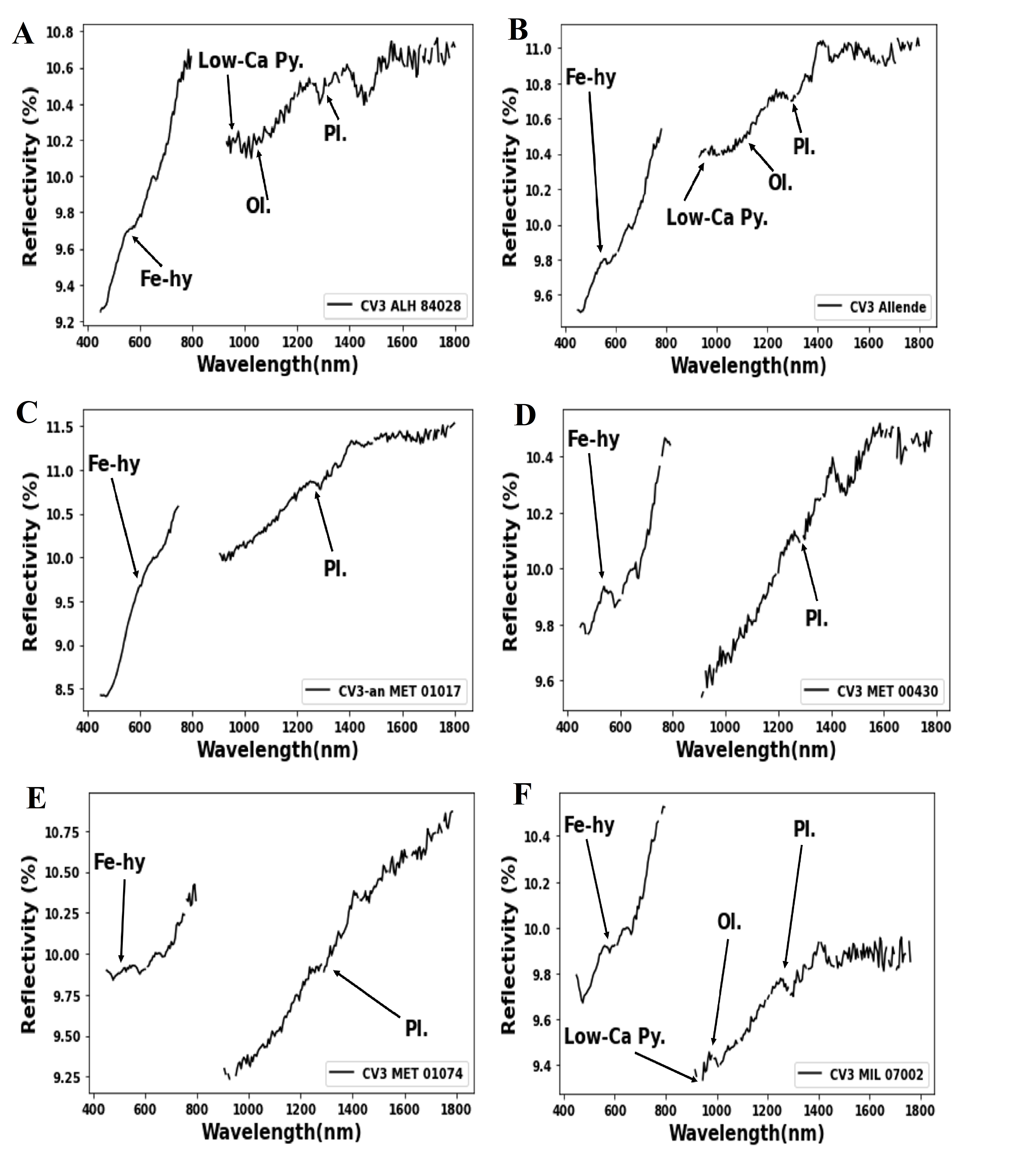}
    \caption{ Reflectance spectra of CV carbonaceous chondrites between 400 and 1900 nm. The spectra were normalized to 8 at 650 nm. Ol. is Olivine, Pl. is Plagioclase feldspear, Fe-hy is Iron-Oxydydroxide and Low-Ca Py. is Low-Ca Pyroxene.}
    \label{fig3}
\end{figure}

The reflectance spectra of CV chondrites, with the exception of Allende, displays a decreasing reflectance toward wavelengths short ward of 500 nm, which is the plausible result of the formation of iron oxyhydroxides due to terrestrial weathering. Significantly, the narrow absorption band in ALH 85002 and Allende, is due to its deep extinction below 400 nm and could be correlated with maghemite and magnetite \citep{tang2003magnetite}. It appears in Fig~\ref{fig3} that fine-grained CVs have shallower olivine absorption bands. Particularly, the 1000 nm olivine absorption band is the main feature observed in CK and CV spectra, but it is deeper in CKs than in CV chondrites.  In our reflectance spectra of CK chondrites, the olivine absorption feature at 1050 nm is dominant, indicating that it is a major rock-forming component. Another distinguishable bands are a low-Ca pyroxene band at 900 nm and a plagioclase feldspar band at 1250 nm (see Table~\ref{tab3}). \citet{cloutis2012spectral} suggests that some CCs do not have well-defined olivine absorption bands, due to not having experienced high enough temperatures to reduce the spectrum-darkening effects of opaques or to recrystallize from pre-existing phyllosilicates.

\begin{figure}

\includegraphics[width=80mm, scale=1.5]{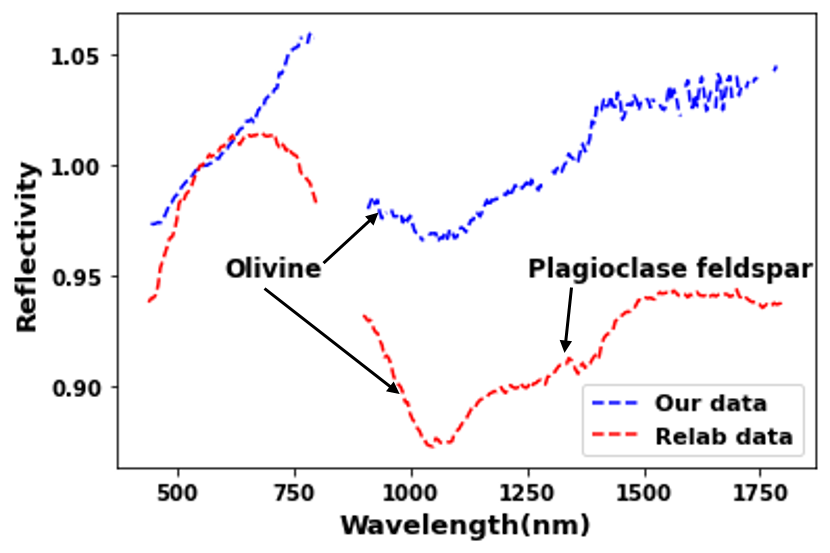}
    \caption{Comparison of the spectra of CK4 chondrite ALH 85002 obtained for this study, and another one from Brown University Keck/NASA RELAB obtained from a powder of the same meteorite. They are normalized to 1 at 550 nm  and a $BaSO_{4}$ correction was applied to ours.}
    \label{fig4}
\end{figure}

Several representative absorption bands have been reported in different CK and CV spectra. First, a narrow band at 600 nm, attributed to FeO and Fe$_{2}$O$_{3}$ contents \citep*{hunt1979spectra} are described in several spectra \citep{lazzarin2004astronomy, belskaya2010puzzling}. There is also a broader band around 650 nm, which is assigned to plagioclase feldspar \citep*{geiger1995formation}. Another commonly found band appears between 900 to 1100 nm, being associated with olivine \citep*{hunt1979spectra}. The most common mineral in CK chondrites is indeed olivine \citep{cloutis2012spectral}. It exhibits an absorption feature centred near 1060 nm that can be seen in our Figure~\ref{fig4}. With increasing Fe$^{2+}$ content, the centre of this absorption band moves to longer wavelengths, becomes deeper and darker by increasing the overall reflectance \citep*[see e.g.][]{king1987relation}. In our data the olivine band extends between 950 and 1150 nm (see Fig~\ref{fig4}), which has shown in Table~\ref{tab2} and Table~\ref{tab3} as the spectral properties of the studied chondrites. Plagioclase feldspar is a minor silicate phase in CK chondrites (< 0.1 wt. \%) that may exhibit a weak Fe absorption band near 1250 nm which appears weakly represented in our spectra \citep*{adams1978plagioclase}.

\begin{figure}
\includegraphics[width=80mm, scale=1.0]{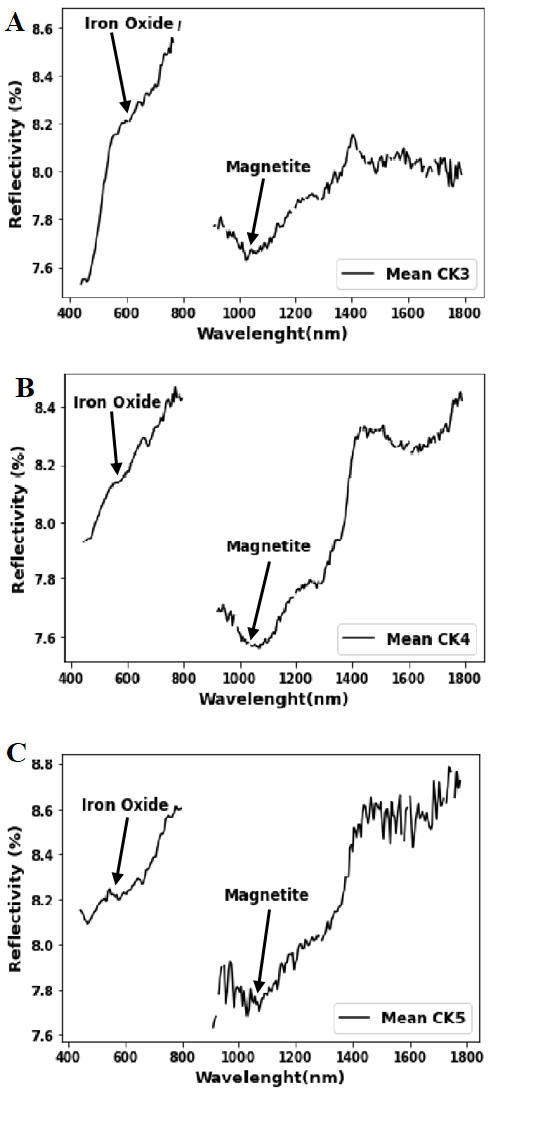}
    \caption{Comparison of reflectance spectra of three petrologic types of CK chondrites in our study (One CK3, two CK4 and two CK5 are mentioned in Table~\ref{tab1}). A $BaSO_{4}$ correction was applied to all.}
    \label{fig5}
\end{figure}

The presence of pyroxene is noticeable as a 2000 nm absorption band in RELAB data, which varies from 1800 to 2080 nm when corresponding to low–Ca pyroxenes, to 1900-2380 nm for high–Ca Pyroxenes \citep{cloutis2012spectral}. Due to the noise introduced by our spectrometer, we cannot properly characterize pyroxene on the samples analyzed here, but the CK spectral data shows a tail of this band near 1850 nm. The presence of pyroxene generally affects the olivine band positions, but this effect is less important in CKs than in other CCs \citep{cloutis1986calibrations}. This is because pyroxene abundances in CKs are generally too low compared to olivine to appreciably affect olivine band positions \citep{cloutis2012spectral}. The spectral shape of the magnetite is a function of grain size and location \citep{morris1985spectral} but it commonly represented by an absorption band that can be seen near 1000 nm. In our CK spectra, the location of this absorption band varies from 980 to 1050 nm depending on the specimen analysed (Fig~\ref{fig5}). We also noticed that other minor phases associated with refractory inclusions can be weakly featured as well, with fassaite, the most common mineral forming Ca- and Al- rich Inclusions (CAIs), being the best example at about 980 nm (Table~\ref{tab2}).

\section{Discussion}
In this section we compare the obtained reflectance spectra with previously published data. This is important, to do this since our procedure obtains the reflectance spectra using polished sections of the meteorites, while previous published results used powders from ground meteorites \citep{trigo2014uv, moyano2017nanoindenting}.

\subsection{Comparison between samples prepared by different methods}

 Here we start the comparison of our reflectance spectra with the most extended procedure that describes the reflectance spectra obtained from meteorite powders. \citet{cloutis2012spectral} identified compositions and spectral features in CK chondrites with small grain sizes (10-30 $\mu$m). The CK chondrites are highly oxidized with petrologic types from 3 to 6, with high abundance of the fayalite Fe$_{2}$SiO$_{4}$ member of the olivine family \citep{geiger1995formation, huber2006siderophile}. The RELAB database at Brown University Keck/NASA also contains CK chondrite spectra covering a wavelength range from 400 to 1900 nm and measured on powders of various sizes \citep{cloutis2012spectral}. Our spectra show significant differences compared with RELAB data shown in Fig~\ref{fig4}. The spectrum of Allan Hills 85002 (ALH 85002) in RELAB has a lower reflectance than the spectrum obtained for this meteorite in our studies. These differences may be due to measuring a polished meteorite surface instead of the fine-grained powders (<125 $\mu$m) which are used in RELAB. \citet{moyano2017nanoindenting} also indicated that the behavior of reflectance spectra in the UV region of the ground samples is remarkably different from our meteorite sections. In our ALH 85002 data the steep slope found by RELAB below 600 nm does not exist. This is possibly produced by terrestrial weathering in the samples studied by RELAB \citep{moyano2017nanoindenting}. \citet*{hendrix2006effects} also mentioned that flattening between 400 and 600 nm indicates high Fe content \citep{cloutis2010reflectance}. As we work with polished samples, the metal grains contribute more to the spectrum than in the powder used in RELAB \citep{trigo2014uv}. As a result, the behavior below 600 nm and other features in our spectra are very similar to changes in space-weathered asteroids \citep*{hendrix2006effects}. It is notable that the narrow absorption band in ALH 85002 around 450 nm (see Fig~\ref{fig1}) could be assigned to magnetite, due to its deep extinction below 400 nm \citep{tang2003magnetite}. We can also notice that the spectrum of ALH 85002 exhibits the lower reflectance in the NIR (see Fig~\ref{fig1}). This is consistent with a CK5 shock-darkened specimen (Table~\ref{tab2}) which has decreased its reflectance due to the presence of fractured materials, melts and opaques. PCA 82500 is another CK chondrite that have experienced shock-darkening so its reflectance spectrum is almost flat in NIR (See  Fig~\ref{fig1}). In consequence, both meteorites are excellent examples of the effect of impact processing in this C-rich asteroids: the initial red slopes are flattened as consequence of shock-darkening to end as neutral in NIR.

To compare with previous papers, \citet{salisbury1975visible} found that the spectrum of CK4 Karoonda, which is a well preserved fall, shows the shallowest absorption features in the 500-600 nm interval and a very weak olivine absorption band in 1000 nm region. \citet{hiroi1993evidence} also identified the blue-sloped reflectance spectra for <125 $\mu$m powders of CK4 ALH 85002 and CK4/5 Yamato 693 (Y 693) in the range of 500-2500 nm, and an olivine-like absorption band near 1050 nm.

\subsection{Comparison of CKs with CV carbonaceous chondrites}

The CK chondrite group was defined by \citet{kallemeyn1991compositional} as having close compositional and textural relationship with the CV chondrites, but they are distinguishable from one another by their refractory lithophile abundances, refractory inclusions abundance, and the presence of igneous rims around chondrules, among other features. As previous author \citet{greenwood2003ck, greenwood2004relationship, greenwood2010relationship} and \citet{devouard2006mineralogy} have suggested, there are some connections between CV and CK chondrites. CKs are also the most oxidized extraterrestrial rocks found so far, owing to their low abundance in Ni and Fe and their high content of fayalite and magnetite \citep*{geiger1995formation}. CV3-an Meteorite Hills 01017 (MET 01017) is possibly a reduced chondrite \citep{busemann2007characterization}. CV3 MET 00430 and CV3 MET 01074 are highly oxidized Bali-like type (CV3$_{OxB}$) \citep{brearley2014heterogeneous, jogo2018redistribution}. CV3 Allende is an oxidized Allende-like type (CV3$_{OxA}$) \citep{bland2000cv3}. CV3  Miller Range 07002 (MIL 07002) is possibly oxidized and intermediate between CV$_{OxA}$ and CV$_{OxB}$ \citep{isa2012bulk}. CV3  Allan Hills 84028 (ALH 84028) is also an oxidized Allende-like type \citep{bland2000cv3} (Table~\ref{tab1}).

\begin{table}
	\centering
	\caption{Spectral features of the spectrum of the meteorites studied here. The estimated error for each value is in the last digit.}
	\label{tab2}
\begin{adjustbox}{center, width=\columnwidth-30pt}
	\begin{tabular}{cccc}
		\hline
		Meteorites & Specific reflectance bands & Band centers & Peak width  \\
		\hline
               
		LAR 04318 & Olivine &	1350 & 0.67  \\
		                & Magnetite &  975	  & 0.67 \\
		                & Iron oxide  & 475 & 3.62 \\
                                 &  Plagioclase felspar  & 1470  & 1.53\\
\hline	

                    PCA 82500 &  Oilivine &	1405 & 3.48 \\
		                & Magnetite & 1025	  & 2.33\\
		                & Iron oxide  & 460 & 1.28 \\
                                  &  Plagioclase felspar  & 1485  & 2.97 \\
\hline

                   ALH 85002 & Magnetite &  1050 &   3.33 \\	
                                                     & Olivine &    1445 & 0.73 \\
		                                          & Plagioclase felspar & 1365	  & 0.82 \\
                                                               & Iron oxide  & 460 & 0.79 \\
\hline

                                    LAR 12265 thick &  Oilivine& 1395 & 0.793 \\
		                                          & Magnetite & 1025  & 0.58\\
                                                        & Iron oxide  & 465 & 5.21 \\
\hline

                                    LAR 12265 thin &  Oilivine&	1315 & 2.41 \\
		                                          & Magnetite & 1010	  &  1.05 \\
                                                               & Iron oxide  & 465 & 4.32 \\
\hline

                                                         ALH 84028 &  Oilivine &	1445 & 0.98\\
		                                          & Low-Ca pyroxene&  1014	  & 2.46 \\
                                                              & Iron oxide & 450 & 0.59 \\
                                                              &  Plagioclase felspar  & 1365  & 0.72\\
\hline

                                    Allende &  Oilivine &	1410 & 1.65 \\
		                                          & Low-Ca pyroxene& 990	  & 1.55\\
                                                              & Iron oxide & 450 & 1.56 \\
                                                        &  Plagioclase felspar  & 1365  & 0.89 \\
\hline

                                           Mill 07002 &  Oilivine &	1300 & 0.58 \\
		                                          & Low-Ca pyroxene& 945	  & 1.82 \\
                                                              & Iron oxide & 465 & 2.28 \\
                                                        &  Plagioclase felspar  & 1335  & 1.58 \\
\hline

                                     MET 01017 &  Oilivine &	1280 & 2.31 \\
                                                   & Low-Ca pyroxene& 915 & 2.10\\
                                                & Iron oxide & 460 & 1.42 \\
                                                        &  Plagioclase felspar  & 1395 & 1.90\\

\hline
                                             MET 00430 & Oilivine &	1225 & 0.74 \\
                                                         & Low-Ca pyroxene& 920 & 1.04\\
                                                  & Iron oxide & 570 & 2.03 \\
                                                        &  Plagioclase felspar  & 1290  & 0.57\\
\hline

                                      MET 01074  &  Oilivine &	1310 &  1.57 \\
                                                         & Low-Ca pyroxene& 960 & 0.91 \\
                                                 &  Iron oxide & 465 & 3.89\\
                                                        &  Plagioclase felspar  & 1395  & 1.25 \\

\hline

	\end{tabular}
\end{adjustbox}
\end{table}

\begin{table}
	\centering
\fontsize{30}{50}\selectfont
	\caption{Comparison between the CK spectral properties of the data in 
 Brown University Keck/NASA RELAB Spectrum PH-D2M-035/C1PH35, and our data, a spectrum taken from the visible and infrared imaging spectrometer at EEBE center of UPC.}
	\label{tab3}
\medskip
\centering
\begin{adjustbox}{center, width=\columnwidth-185pt}
	\begin{tabular}{cccc}
		
\hline
	Wavelengths (nm)  &  Wavelengths (nm)    & Wavelengths (nm)  & Minerals \\
                                         (Ours)	&  \citep{cloutis2012spectral}	& (RELAB)  &    \\	 
\hline
		
	580    & 650    &    &  Iron oxide (FeO) \\

	980  & 950-1100   &  1100-1200 &   Magnetite (Fe$_{3}$O$_{4}$)  \\
 
	1380   &  1250   &   1050   &     Olivine ((Mg,Fe)$_{2}$SiO$_{4}$)  \\

940 & -  & - &  Low-Ca pyroxene \\

	1300	    	&	1250	&    1350     &     Plagioclase Feldspar \\

	2100	  & 2100	 &            &     Fassaite \\
\hline			
	\end{tabular}
\end{adjustbox}
\end{table}

\begin{table}
	\centering
	\caption{Average spectral slope and the reflectivity of the meteorites at some specific points. First coloumn is the name of the meteorite sample, the second coloumn is the Average spectral slope (\%/ nm) in the indicated range, the third, forth and last coloumn are the reflectivity at designated wavelengths. The estimated error for each value is in the last digit.}
	\label{tab4}
\begin{adjustbox}{center, width=\columnwidth-50pt}
	\begin{tabular}{ccccc}
		\hline
		Met. & Av. in (1000-1800 nm)  & Ref. at 600 nm & Ref. at 1200 nm  & Ref. at 1800 nm  \\

		\hline

		LAR 04318 &  0.0016 & 9.90 & 8.00 & 8.90 \\

                               PCA 82500  & 0.0004  &  9.05 & 8.68 & 8.80 \\

                                ALH 85002 & 0.0062  & 8.19 & 8.00 & 8.41\\

                                    LAR 12265 thick  & 0.0014  & 8.21 & 7.90 & 8.70 \\

                                    LAR 12265 thin & 0.0013 & 8.08 & 7.90 & 8.70\\

                                    ALH 84028 & 0.0065  & 9.70 & 10.48 & 10.7 \\

                                    Allende & 0.0007 &  9.81 &  10.70 & 11.00 \\

                                           Mill 07002 &  0.0016  & 9.90 & 9.65 & 9.90 \\

                                     MET 01017 & 0.0001 & 9.75 & 10.75 & 11.5 \\

                                             MET 00430 & 0.0019  & 9.90 & 9.95 & 10.50 \\

                                      MET 01074  & 0.0006 & 9.85 & 9.75 & 10.90 \\
		
	\end{tabular}
\end{adjustbox}
\end{table}

\begin{figure}
	\includegraphics[width=90mm, scale=1.5]{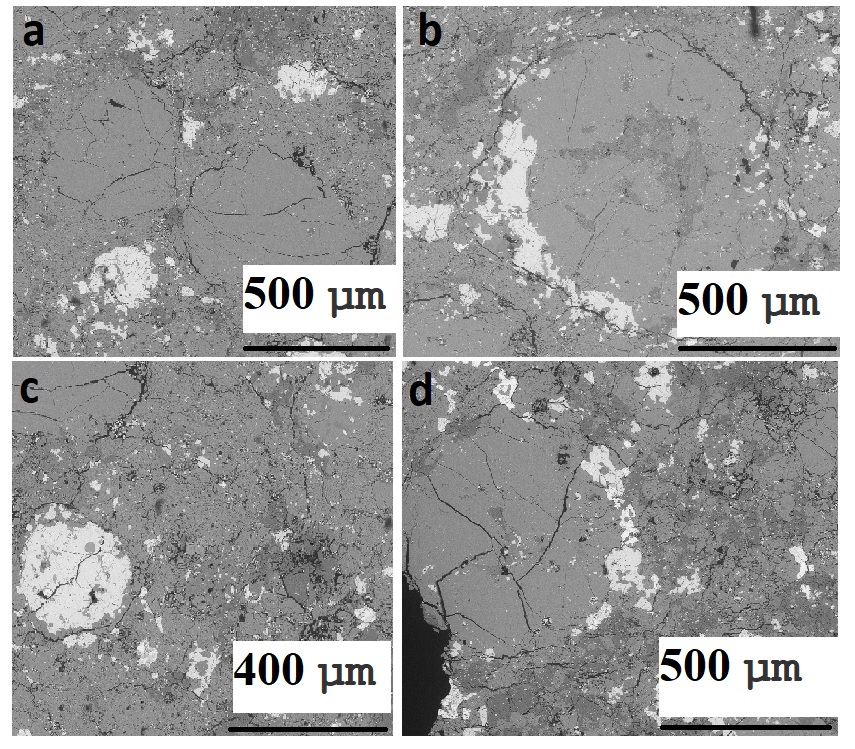}
    \caption{Back-scattered electron (BSE) image of regions of CK4 chondrite
ALH 85002. It is a nice example of well preserved CK consisting of well
compacted silicate-rich chondrules, sulfides and accessory minerals.
The dominant minerals and their respective tonality are: olivine – light
gray; orthopyroxene – medium gray; plagioclase – dark gray; sulfide
and magnetite – white; fractures and pores being black.
}
    \label{fig6}
\end{figure}

The importance of collisional gardening of the CK parent asteroid was proposed because of different features noticed in their matrices \citep*{rubin2017meteoritic}. Matrix-rich regions in the CK chondrites formed by comminuted and ground materials, mostly micron-sized angular silicate grains that are pressed and crushed chondrules, demonstrate the degree of collisional processing in these samples \citep*{rubin2017meteoritic}. Back-scattered electron (BSE) images of ALH 85002 exemplify the internal fractures usually contained in their rock-forming silicates and the subsequent comminution of the materials. As consequence of these processes secondary minerals were formed, acting as opaques that change the spectral reflectance behavior (Fig~\ref{fig6}). On the other hand, the primary mineralogy of CVs seem to be better preserved than the CKs, so seems plausible that they suffered a minor extent of collisional processing. For example, thin sections of CV3$_{oxA}$ Allende \citep{norton2002book} reveal little evidence of crushing. Although many reduced CV3 chondrites show impact signs \citep{scott1992shock}, they have not been extensively crushed \citep{kracher1985leoville}.

\subsection{Comparing laboratory spectra of CCs with remote asteroidal reflectance spectra}

As a result of the reflectance affinity and the chemical evidence found by \citep*{wasson2013compositional}, we found plausible that CV and CK chondrites were associated with a common parent asteroid that experienced splitting and differentiated evolution by consequence of separated orbital evolution. \citet{greenwood2010relationship} and \citet{cloutis2012spectral} suggested that CVs and CKs may have a common parent body,  Eos asteroid family should be its most likely source. It is also consistent with the suggestion that the 221 Eos asteroid collisional family proposed to be associated with CV meteorites \citep{greenwood2010relationship}, could be a possible source of CK chondrites since they share a a blue-sloped features in the NIR region (800-2500 nm) specially beyond $\sim$1500 nm \citep{cloutis2012spectral}. Then, collisionally processed fragments of 221 Eos family could be the source of high petrologic chondrites.

\begin{figure}
	\includegraphics[width=75mm, scale=0.6]{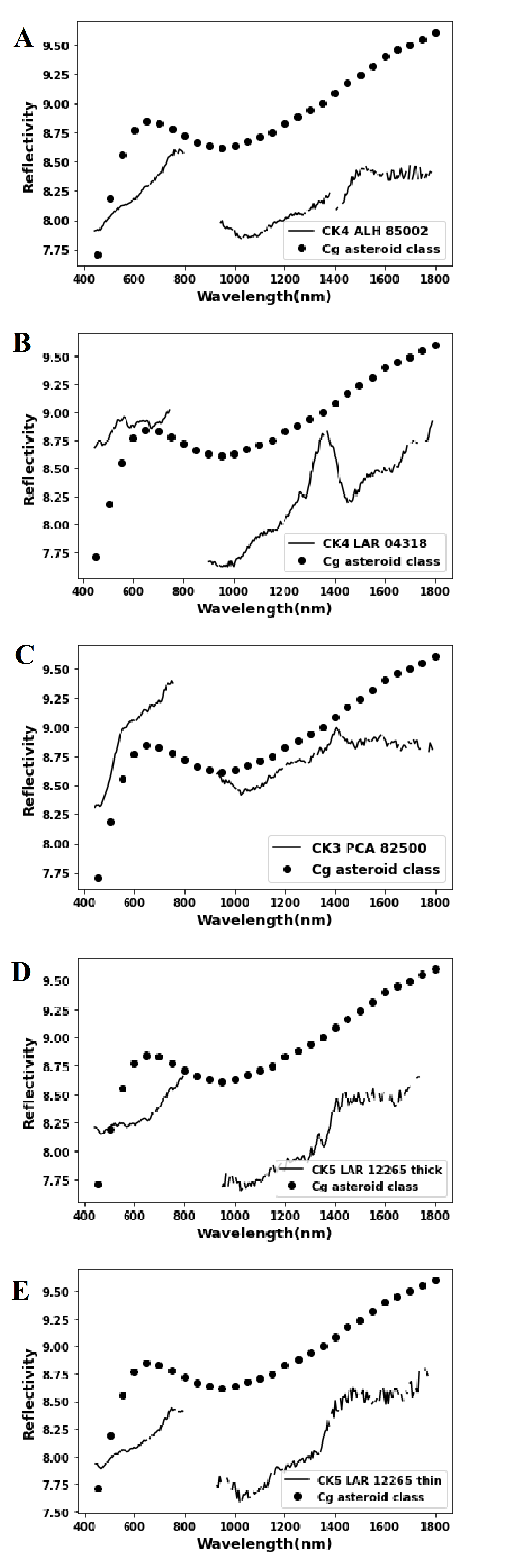}
    \caption{CK reflectance spectra compared with reflectivity of Cg-type asteroids in the UV-NIR range. The Cg spectrum is normalized and the multiplied with a reflectance value in order to bring it on the same scale.}
    \label{fig7}
\end{figure}

 We have compared our mean reflectance spectra with the Cg asteroidal reflectance class of the Bus-DeMeo Taxonomy Classification \citep{demeo2009extension}. It is important to remark that \citet{mothe2005221} and \citet{mothe2008mineralogical} compared spectra of 221 Eos asteroid members with the spectra of different CC groups. They previously realized in that work that many members of that asteroid family are consistent with CK3, CK4 and CK5 chondrites in the NIR region, but differ in the visible region \citep{salisbury1975visible}. 221 Eos and its family members are classified as K asteroids in the Bus-DeMeo taxonomy \citep{demeo2009extension} (See Fig~\ref{Fig2a} in the appendix) . We used a Cg class spectrum for comparison with the spectra of CK chondrites in the range of 400-1800 nm (Fig~\ref{fig7}).


Moreover, the recent simulated space weathering experiments on CV3 meteorite Allende by \citet{gillis2017incremental} showed that the weathered meteorite shows a different weathering trend for the continuum slope between 450-550 nm. The absorption band in 900-1400 nm region in CV3 meteorites resembles the same region in Ryugu although it is not well-defined there \citep{de2018expected}. The absorption bands in the range of 1050 to 1250 nm in ureilites corresponds to olivine, so we infer that the same bands  in CVs and CKs are also associated with olivine \citep{cloutis2010reflectance, cloutis2012spectral, le2017ground}.

Our results show that Cg spectral type asteroids could match the spectra of CK chondrites according to what was found between 500 and 800 nm (Fig~\ref{fig7}).  In general, several specific features in an asteroid reflectance spectrum display the effect of aqueous alteration in the asteroid surface \citep{barucci2005asteroid}. As we combined our spectra with the Bus-DeMeo Taxonomy Classification tool, it is realized that the overall shape of most CK spectra seems to be similar to Cg-type asteroids \citep{demeo2009extension}. 

Recent analytical work on the CV and CK magnetite composition, found differences in Cr isotopic abundances \citep{dunn2016magnetite, yin2017testing}. Nevertheless, the parent bodies could be very similar, because those
findings on minor elements cannot account for significant differences in their reflectance spectroscopic characteristics. Our working hypothesis suggesting both CV and CK groups could be the result of a differentiated outcome from separate bodies belonging to a collisional family could explain those subtile differences. In addition, such scenario could be consistent with the presence of two
CV7 chondrites listed in the Meteoritical Bulletin. The presence of such metamorphosed CVs could suggest two separate metamorphic sequences, but they could be rare impact melts and not type-7 after all. Then, it seems that more analytical data is needed to demonstrate that CV and CKs are not related as we propose.
\subsection{Eos family comparisons}

\begin{figure}
	\includegraphics[width=80mm, scale=1.5]{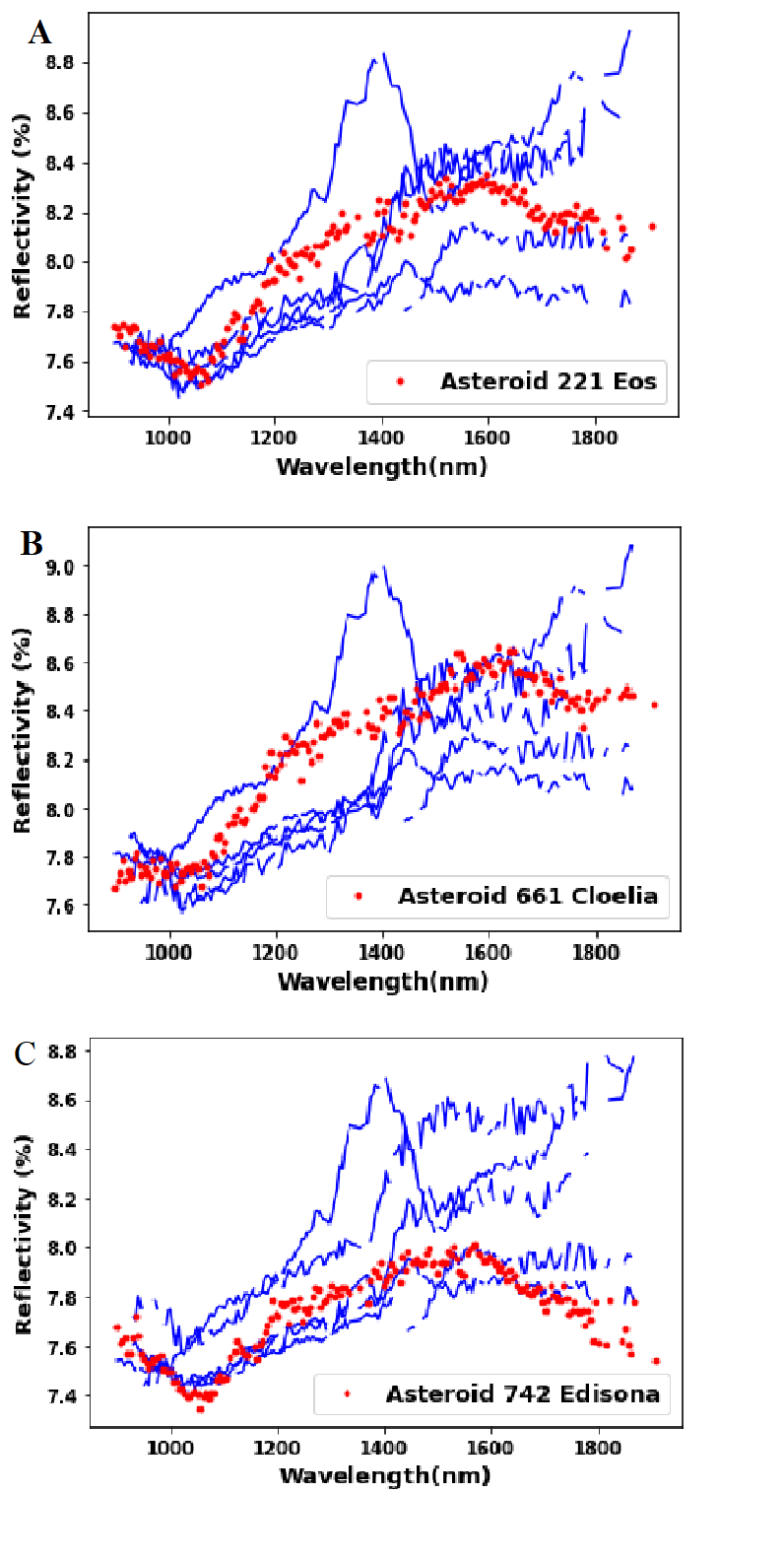}

    \caption{CKs reflectance spectra compared with reflectivity of A) 221 Eos, B) 661 Cloelia and C) 742 Edisona in the range between 900 nm to 1900 nm. Blue ones are the spectra of ALH 85002, LAR 04318, PCA 82500, LAR 12265 thick, LAR 12265 thin which are shown in the previous plots.}
 
    \label{fig8}
\end{figure}

\begin{figure}
	\includegraphics[width=80mm, scale=1.5]{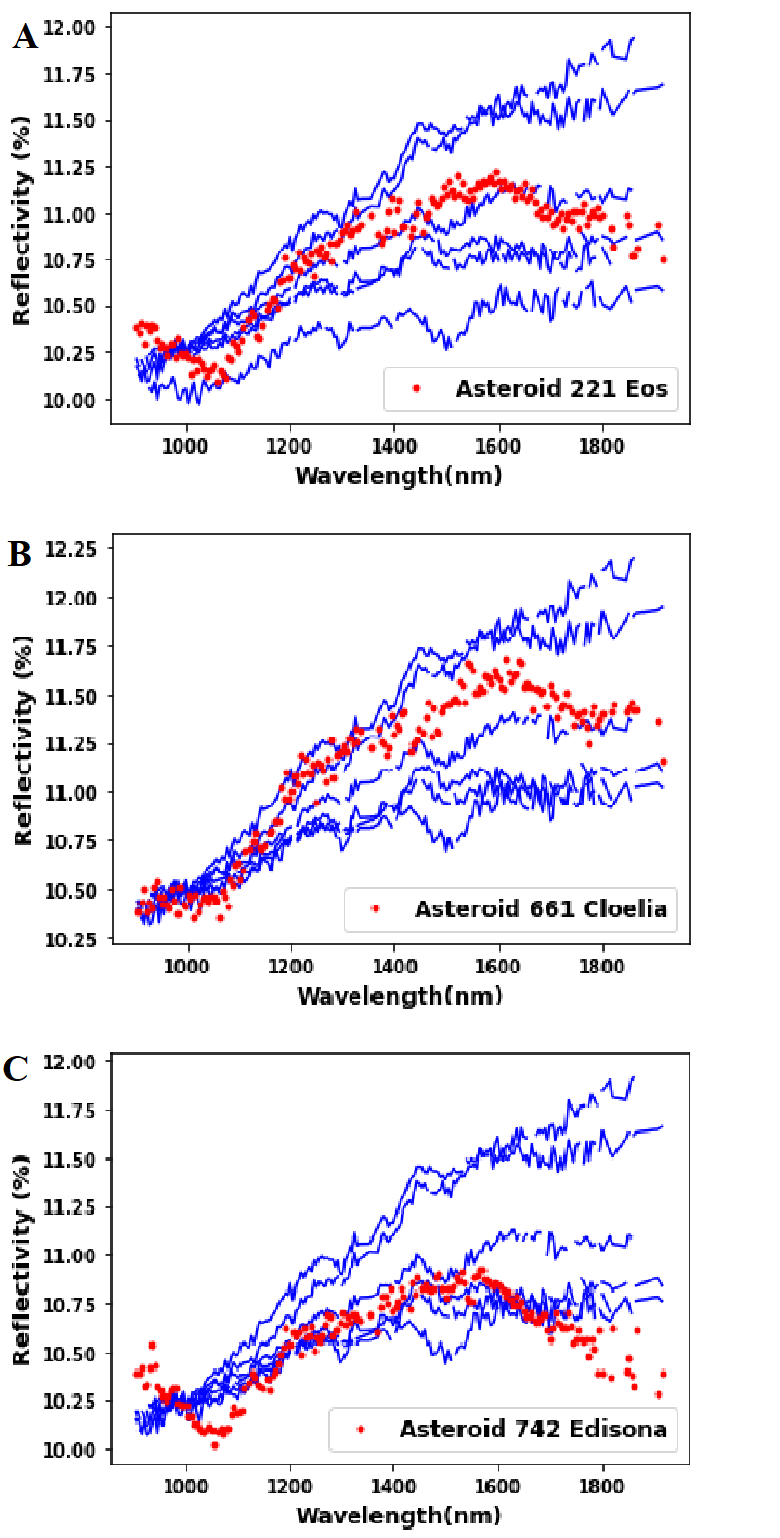}
	
    \caption{CVs reflectance spectra compared with reflectivity of A) 221 Eos, B) 661 Cloelia and C) 742 Edisona in the range between 900 nm to 1900 nm. Blue ones are the spectra of ALH 84028, MIL 07002, Allende, MET 00430, MET 01017, which are shown in the previous plots.}
 
    \label{fig9}
\end{figure}

In order to demonstrate that the CV and CK groups could be a good match for the Eos family we have compared three asteroids of that family: 221 Eos, 661 Cloelia and 742 Edisona with our meteorite spectra \citep{clark2009spectroscopy}. Fig.~\ref{fig8} shows the spectral comparison between CKs and the three asteroids, while the CVs are compared in Fig.~\ref{fig9}. The spectral slope of the asteroids match well these exhibited by the CV and CKs. In addition, there are some common noticeable features. The spectra of 221 Eos and 661 Cloelia exhibit two olivine bands at 1050 and 1450 nm that also noticeable in our CK spectra, with exception of the anomalous LAR 04318. Also the minor band associated with plagioclase feldspar at 1370 nm is also a common feature between the three asteroids and our CKs.

 In the NIR the reflectance spectrum of 221 Eos has resemblance with the flat behavior of ALH 85002, while such flat behavior is more extreme for asteroid 742 Edisona that exhibits a NIR spectrum resembling that of the shock-darkened PCA 82500. Concerning the CV reflectance spectra, they are having a similar slope, but the olivine features are not so marked as these exhibited by the three asteroid spectra. \citet{morate2018color} found a wide diversity of near-infrared colours for 328 members of this family which highlights the compositional variation among them.

\section{Conclusions}

We have tested a working hypothesis for the formation of the parent bodies of CV and CK chondrites. The CKs could have formed from CVs after impacts and high temperatures, and these processes made them aqueously altered and annealed \citep*{wasson2013compositional}. Such a working scenario implies an evolutionary processing of CKs from CV materials and suggests that meteorite collections should have specimens exhibiting gradational differences with significant consequences in their reflectance spectra. Such scenario is consistent with our current study of CV and CK reflectance spectra. Then, from the study of CV and CK reflectance spectra we reach the following conclusions:

 \begin{enumerate}
\item We found a significant similitude between the reflectance spectra of CV-CK chondrites and asteroids belonging to the Cg spectral class. Specific diversity could be consequence of thermal heating due to different degree of collisional processing. 

\item A common behavior for CK chondrite reflectance spectra is found. We realize that with increasing metamorphic grade, the reflectance decreases producing a deeper 1000 nm region absorption feature. But in general, higher petrographic degree is consistent with higher reflectivity, mainly in the visible region. The band depth of olivine in CKs increases while the thermal metamorphism increases: the CK3 petrographic grades of meteorites display shallower band depths, while the CK5 spectra exhibit the deepest band depths. 

\item The correlation between CV and CK reflectance spectra, together with the chemical similitude claimed in previous studies, suggest that both groups could have originated from a common parent body that was broken apart a long time ago. In this evolutionary scenario, different asteroid fragments could have been dynamically separated by the action of non-gravitational forces. As a consequence of differentiated collisional evolution, each asteroid ended with different degrees of space weathering, aqueous alteration and thermal metamorphism.

\item Collisional processing is also clearly affecting these meteorites. As consequence of shock, the main silicates are crushed and fragmented, producing shock-induced minerals and opaques that remain in the minerals matrixes. The main consequence is that the overall reflectance decreases and the spectra change from red slope to a more neutral behavior in the NIR region.

\item If our evolutionary scenario is correct, secondary minerals could be formed as a natural consequence of the collisional processing and aqueous alteration of their parent asteroids. A natural increasing in the opaque minerals in the matrices of these meteorites could have direct influence on the progressive darkening found in the spectra of CV-CK chondrites.

 \end{enumerate}

\section*{Acknowledgements}

This research has been funded by the research project (PGC2018-097374-B-I00, P.I.: JMT-R), funded by FEDER/Ministerio de Ciencia e Innovación – Agencia Estatal de Investigación. ST is hired in the framework of that research project. This study was done in the frame of a PhD. on Physics at the Autonomous University of Barcelona (UAB). JL is a Serra Húnter Fellow and is grateful to ICREA Academia program and GC 2017 SGR 128. US Antarctic meteorite samples are recovered by the Antarctic Search for Meteorites (ANSMET) program which has been funded by NSF and NASA, and characterized and curated by the Department of Mineral Sciences of
the Smithsonian Institution and Astromaterials Acquisition and Curation Office at NASA Johnson Space
Center. The Institute of Space Sciences (CSIC) is international repository of these Antarctic
samples. We thank all these institutions for providing the Antarctic meteorites studied here. This research utilizes spectra acquired by with the NASA RELAB facility at Brown University.


\section*{Data availability}

Spectral data in this paper are subjected to one year embargo from the publication
date of the article to allow the completion of the research project and conclude the meteorite spectral database. Once the embargo expires the data will be available in a repository upon reasonable request.


\bibliographystyle{mnras}
\bibliography{ref} 


\clearpage
\appendix

\section{Complementary figures}

\begin{figure}
	\includegraphics[width=\columnwidth]{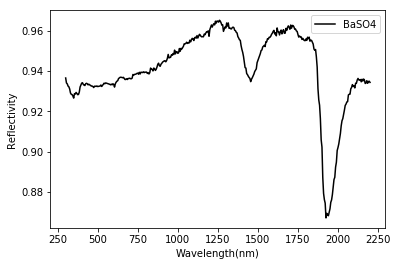}
    \caption{Barium sulphate ($BaSO_{4}$) standard baseline, used in the correction applied to calibrate the detector \citep{moyano2016plausible}.}
    \label{Fig1}
\end{figure}

\begin{figure}
	\includegraphics[width=80mm, scale=0.9]{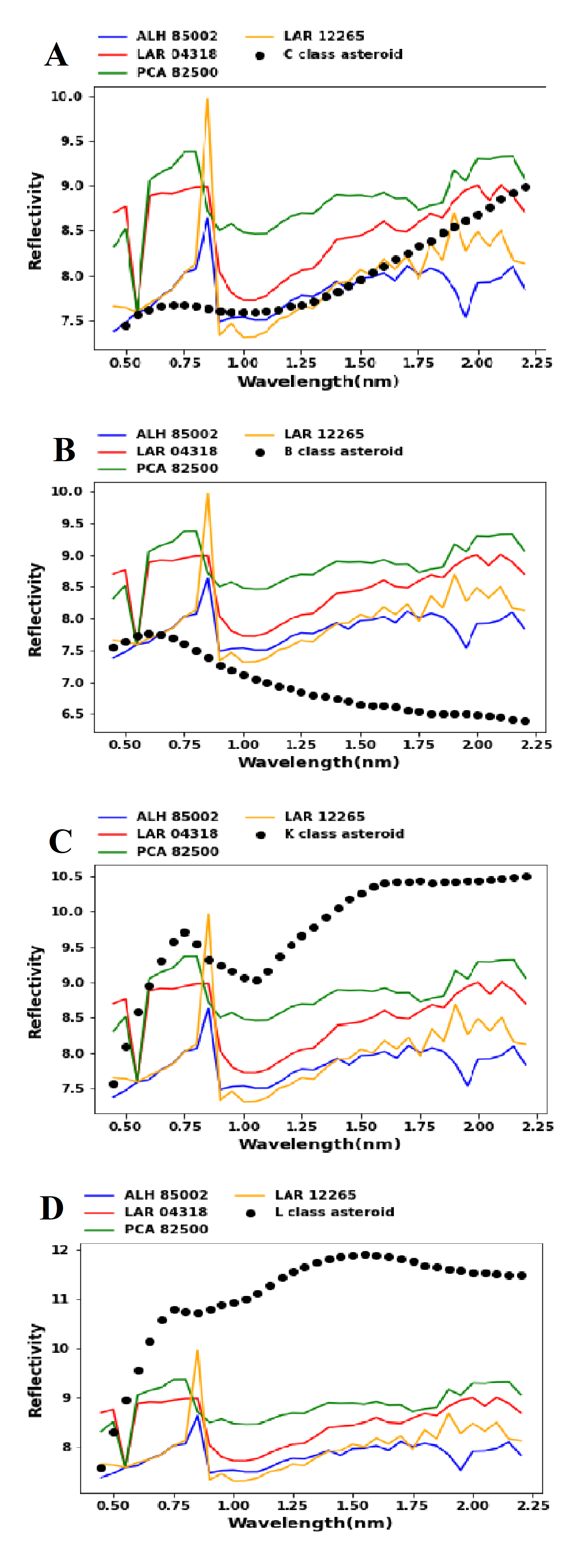}
    \caption{ Spectral comparisons of CK chondrites with other averaged asteroid
spectral classes ruled out during the investigation. A)C ateroid classes, B) B asteroid classess, C)K asteroid classes and D)L asteroid classes}.
    \label{Fig2a}
\end{figure}

\clearpage 

\section{Python Code}

\citet{10.5555/1593511,van1995python}

import pandas as pd\\
import matplotlib.pyplot as plt\\
from scipy.signal import (find$\_$peaks), peak$\_$widths\\
import numpy as np\\

\citet{2020SciPy-NMeth, 2020NumPy-Array, 4160265, mckinney-proc-scipy-2010}

\# get max peaks \\
peaksMax, $\_$= find$\_$peaks(CK4$\_$LAR$\_$04318, distance=70) \\

\# get min peaks \\
peaksMin,$\_$= find$\_$ peaks(-CK4$\_$LAR$\_$04318, distance=70) \\

\# get max peak width \\
widths, h$\_$eval, left$\_$ips, right$\_$ips = peak$\_$widths(-CK4$\_$LAR$\_$04318, peaksMax) \\

\# get min peak width \\
widths, h$\_$eval, left$\_$ips, right$\_$ips = peak$\_$widths(-CK4$\_$LAR$\_$04318, peaksMin) \\
print("Widths of each min peak: ", widths) \\






---------------------------------------------------------------------------------------------------

\bsp	
\label{lastpage}

\end{document}